\documentclass[prb,showpacs,twocolumn, amsmath,amssymb]{revtex4}

\usepackage{graphicx}
\usepackage{dcolumn}
\usepackage{bm}
\newdimen\figurewidth
\begin{document}
\preprint{APS/123-QED}

\title{Numerical study of the Coulomb blockade in an open quantum dot}

\author{Yuji Hamamoto}
\email{hamamoto@issp.u-tokyo.ac.jp}

\author{Takeo Kato}
\affiliation{%
Institute for Solid State Physics, University of Tokyo, Kashiwa,
Chiba 277-8581, Japan
}%


\date{\today}

\begin{abstract}
The Coulomb blockade in an open quantum dot connected to a bulk lead
by a single mode point contact is studied numerically
using the path-integral Monte Carlo method.
The Coulomb oscillation of the average charge and capacitance of the dot
is investigated,
and is compared with the analytic expression for strong tunneling.
At the degeneracy point, we observe logarithmic divergence of the capacitance
for strong backscattering at the point contact. This observation
supports the conjecture that the nature of the present system
at the degeneracy point is described by the two-channel Kondo problem
for an {\it arbitrary} strength of tunneling.
\end{abstract}











\maketitle

In a mesoscopic structure called a quantum dot,
interaction between electrons restrict
the amount of the inner charge $Q$ at low temperatures $T\ll E_C$,
where $E_C=e^2/2C_0$ is the charging energy, and 
$C_0$ is the static capacitance of the dot.
The electrostatic energy of the dot is given as
$E_Q = (Q-eN)^2/2C_0$,
where $N$ is proportional to the gate voltage $V_g$.
If transport of electrons is governed by weak electron tunneling
between leads and a dot, one observes sharp periodic peaks
of conductance as a function of the gate voltage. Here, the peaks
correspond to the degeneracy points $N=n+1/2$ between two charge
states, $Q=ne$ and $Q=(n+1)e$. Away from the peaks, the amount of 
the inner charge is fixed as $Q=ne$, where $n$ is an integer minimizing $E_Q$.
These phenomena are called the Coulomb blockade. 
For strong tunneling, which is realized in a quantum dot made in semiconductor
heterostructures, the Coulomb blockage is suppressed by large charge fluctuation in the dot.

In order to deal with the strong tunneling regime, it is convenient to
consider a system consisting of a dot and a half-infinite lead
connected by a single mode point contact.
The Coulomb blockade in such a structure can be observed
by measuring the average charge and the capacitance,
and have been studied both 
theoretically\cite{matveev1,matveev2,lehur,lebanon}
and experimentally.\cite{berman}
The peculiar nature of this system is logarithmic divergence of the 
capacitance characterized by the analogy to
the two-channel Kondo problem\cite{cox};
Matveev has revealed that in the present system logarithmic divergence 
of the capacitance appears at the degeneracy points in the regime of
both weak and strong tunneling through the point contact.\cite{matveev1,matveev2}
The consistency in these opposite limits suggests that
the Coulomb blockade system at the degeneracy points can be described effectively by
the two-channel Kondo model for an {\it arbitrary} strength of tunneling.
One way of confirming this conjecture is to
numerically study the intermediate tunneling regime.
Recently, the numerical renormalization group (NRG) method has been applied to
this problem,\cite{lebanon}
and nonmonotonic growth of the Coulomb blockade has been reported
in the strong tunneling regime.
However, this calculation is based on a model of the weak tunneling regime.
It is quite necessary to study this problem by using a model applicable
to both weak and strong tunneling.

In this paper, we study the Coulomb blockade phenomena for an intermediate tunneling
in an open quantum dot---a dot strongly connected to a bulk lead.
If the curvature of the constriction near the center of the point contact
is smooth, the system is essentially one-dimensional, and described by 
the Tomonaga--Luttinger liquid (TLL): an effective model of low energy excitations
in an interacting one-dimensional electron system.
According to Ref.~\onlinecite{matveev2},
the Hamiltonian of the system is bosonized as
\begin{align}
H=&\sum_{\nu=\rho,\sigma}\int\frac{dx}{4\pi}
\biggl[\frac{u_\nu}{K_\nu}
\biggl(\frac{\partial\phi_\nu}{\partial x}\biggr)^2
+u_\nu K_\nu\biggl(\frac{\partial\theta_\nu}{\partial x}\biggr)^2
\biggr]\nonumber\\
&+U[\phi_\rho(x=0)-\pi N]^2\nonumber\\
&+V\cos\phi_\rho(x=0)\cos\phi_\sigma(x=0),\label{ham}
\end{align}
where we assume that the dot (lead) corresponds to the region $x>0$ ($x<0$).
The subscripts $\rho$ and $\sigma$ stand for
a charge and spin mode, respectively.
The first term in r.h.s. of Eq.~(\ref{ham}) describes a bulk spinful TLL
with the sound velocity $u_\nu$.
The positive parameter $K_\nu$ represents an interaction strength:
$K_\nu<1$ for the repulsive case; $K_\nu>1$ for the attractive case;
$K_\nu=1$ for the noninteracting case.
Charging effect in the dot
and backscattering of electrons at the point contact
correspond to the other two terms in Eq. (\ref{ham}),
whose coefficients are defined as 
\begin{gather}
U\equiv\frac{1}{\pi^2}\frac{e^2}{2C_0}=\frac{1}{\pi^2}E_C,\quad V\equiv\frac{2rD}{\pi}.
\end{gather}
Here $r$ is the reflection amplitude, $D\equiv 2\pi/\Delta\tau$ is the bandwidth cutoff,
and $N\equiv C_0V_g/e$ is the dimensionless gate voltage. We note that the present
model can cover the whole region between the weak tunneling limit ($r \rightarrow 0$) and the
strong tunneling limit ($r \rightarrow 1$).

Since the charging and backscattering term in Eq.~(\ref{ham}) are expressed
only by the fields at the origin,
we can integrate out the boson fields in the bulk part and obtain the effective action
\begin{align}
S&\equiv S_0+S_C+S_V,\label{act}\\
S_0&=\sum_{\nu=\rho,\sigma}\sum_{\omega_n}\frac{|\omega_n|}{2\pi K_\nu\beta}
|\tilde{\phi}_\nu(\omega_n)|^2,\\
S_C&=U\int d\tau\,[\phi_\rho(\tau)-\pi N]^2,\\
S_V&=V\int d\tau\cos\phi_\rho(\tau)\cos\phi_\sigma(\tau),
\end{align}
where the spatial coordinate $x$ is omitted,
and $\tilde{\phi}_\nu$ denotes the Fourier component of $\phi_\nu$.
The path-integral Monte Carlo (PIMC) method\cite{herrero,werner} is one of the powerful ways
to simulate quantum systems described by an effective action such as Eq.~(\ref{act}).
Recent development\cite{werner} of this method has remarkably improved
the efficiency of the simulation at low temperatures.
Our Monte Carlo simulation is constructed by
local update in the Fourier space and
global update based on the cluster algorithm.\cite{werner,swendsen}
Similar simulation has been performed in our previous work,\cite{hamamoto}
which has discussed the impurity problem in a spinful TLL corresponding to the case of $U=0$.

Discretizing the imaginary time into $L$ steps,
we define the $j$th step on a path as
$\phi_{\nu j}\equiv\phi_{\nu}(j\beta/L)$, and its Fourier transform as
$\tilde{\phi}_{\nu k}\equiv\sum_j\phi_{\nu j}{\rm e}^{(2\pi i/L)jk}
-\pi LN\delta_{\nu\rho}\delta_{k 0}$.
We thus obtain the discretized effective action
suitable to the local update as
\begin{align}
S_0+S_C&=\sum_{\nu=\rho,\sigma}\sum_{k=0}^{L/2}\frac{1}{2\sigma_{\nu k}{}^2}
|\tilde{\phi}_{\nu k}|^2,\\
S_V&=V\Delta\tau\!\sum_{j=0}^{L-1}\cos\phi_{\rho j}\cos\phi_{\sigma j},
\end{align}
where $\Delta \tau = \beta/L$.
In a local update, for each pair of $\nu$ and $k$ a value of $\tilde{\phi}_{\nu k}$
is randomly chosen following a normal distribution
$\propto {\rm e}^{-|\tilde{\phi}_{\nu k}|^2/(2\sigma_{\nu k}{}^2)}$
with variance~\footnote{
The variance $\sigma_{\sigma 0}=\infty$ means that
a {\it uniform} distribution, e.g., ranged from $-\pi$ to $\pi$ is used
for a local update of $\tilde{\phi}_{\sigma 0}$.
}
\begin{align}
\sigma_{\rho 0}{}^2&=\frac{L}{2U\Delta\tau},\quad
\sigma_{\sigma 0}{}^2=\infty,\\
\sigma_{\nu L/2}{}^2&=\frac{1}{2}\biggl[\frac{1}{2K_\nu L}
+\frac{U\Delta\tau}{L}\delta_{\nu\rho}\biggr]^{-1},\\
\sigma_{\nu k}{}^2&=\frac{1}{4}\biggl[\frac{k}{K_\nu L^2}
+\frac{U\Delta\tau}{L}\delta_{\nu\rho}\biggr]^{-1}\quad (k\ne 0,L/2).
\end{align}
Then a new path $\phi_{\nu j}$ is obtained
from the inverse Fourier transform, 
and is accepted with probability $p={\rm min}\{1,{\rm e}^{-\Delta S_V}\}$.
The local update in the Fourier space globally changes the paths in the real space.
In the presence of strong backscattering, however,
relevant update tends to be rejected at lower temperatures, because
such global changes do not consider the potential configuration.

The basic idea of the cluster algorithm is as follows.
A reflection mirror is suitably located, and
every two sites in a discrete path are connected with
a certain probability.
Then reflection of the connected sites against the mirror
is accepted with a probability determined by
the cost of reflection of the cluster.
If we rewrite the effective action as
\begin{align}
S_0&=-\sum_\nu\sum_{j<j'}\kappa_{\nu jj'}\phi_{\nu j}\phi_{\nu j'},
\label{long}\\
S_C&=U\Delta\tau\sum_j[\phi_{\rho j}-\pi N]^2,\label{charging}\\
S_V&=V\Delta\tau\sum_{j}\cos\phi_{\rho j}\cos\phi_{\sigma j},\label{reflection}
\end{align}
with the kernel defined by
\begin{gather}
\kappa_{\nu jj'}=-\frac{2}{K_\nu L^2}\sum_{k=-L/2+1}^{L/2}
|k|{\rm e}^{(2\pi i/L)(j-j')k},
\end{gather}
we can design two types of cluster update as follows.
In a {\it single}-field cluster update,
clusters for the charge and spin field are constructed separately with
the bond probability between the $j$th and $j'$th site
\begin{gather}
p_{\nu jj'}={\rm max}\{0,1-\exp({-2\kappa_{\nu jj'}
\varphi_{\nu j}\varphi_{\nu j'}})\}.
\end{gather}
Here we define the relative field $\varphi_{\nu j}\equiv\phi_{\nu j}-\pi M_\nu$
measured from the mirror located at $\phi_\nu=\pi M_\nu$ with an
integer $M_\nu$.
In a {\it double}-field cluster update, on the other hand,
a pair of the $j$th and $j'$th site for both the charge field
and the spin field is
connected using the probability
\begin{gather}
p_{jj'}\!=\!{\rm max}\biggl\{0,1-\exp\biggl(-2\sum_\nu\kappa_{\nu jj'}
\varphi_{\nu j}\varphi_{\nu j'}\biggr)\biggr\},
\end{gather}
with a half integer $M_\nu$.
Clearly, $S_V$ remains unchanged after these global moves,
and the strength of backscattering does not matter.
$S_C$ is invariant exceptionally in the single-field case for the spin mode,
but it generally changes in the other cases.
Then the cluster updates for the latter cases are accepted
with the probability $p={\rm min}\{1,{\rm e}^{-\Delta S_C}\}$,
where the change in $S_C$ has the form
\begin{gather}
\Delta S_C=4\pi U\Delta\tau(N-M_\rho)\sum_{j\in{\rm cluster}}
\varphi_{\rho j}.
\end{gather}
The kinks, which run from a potential minimum to an adjacent one
of the double-cosine potential in Eq.~(\ref{reflection}), are
inserted in the path by these two types of cluster updates.
These kinks are relevant structures when one considers strong
backscattering (corresponding to weak tunneling) at low temperatures.

\begin{figure}[t]
\begin{center}
\includegraphics[width=.9\figurewidth]{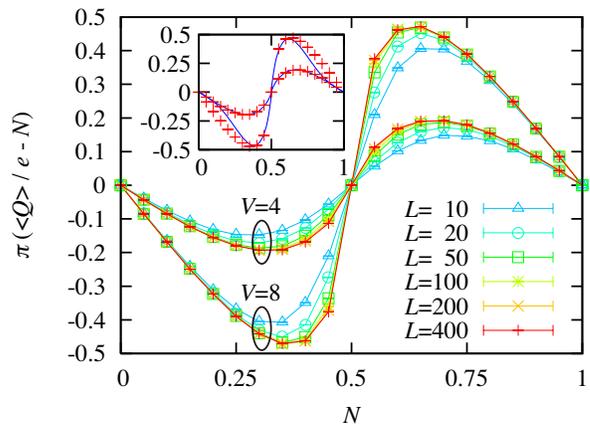}
\caption{\label{fig_q_u1}(Color online) Coulomb oscillation of the average charge
of the quantum dot for different temperatures $L=10$, $20$, $50$, $100$,
$200$, and $400$.
A bundle of curves with smaller amplitude corresponds to $V=4$,
and the other to $V=8$.
Inset:
Comparison of the average charge for $V=4$ and $8$
between our PIMC data at $L= 400$ (crosses)
and the analytic result for small $V$ at zero temperature
(solid line).
}
\end{center}
\end{figure}

In what follows, the results of our PIMC simulation are presented.
We use $U=1$ as the energy unit,
and fix the width of the time step to $\Delta\tau=1/4$.
A noninteracting TLL ($K_\rho=K_\sigma=1$) is assumed in the bulk part.
Because of the periodicity of the double-cosine potential in $S_V$
[see Eq. (\ref{reflection})], 
we can restrict the value of $N$ in the range $0\le N\le 1$.
Since the charge in the dot is bosonized as
$Q\equiv(e/\pi)\phi_\rho(x=0)$,
the average charge is measured as
\begin{gather}
\langle Q\rangle=\frac{e}{\pi}\langle\bar{\phi}_\rho\rangle,\quad
\bar{\phi}_\rho\equiv\frac{1}{\beta}\int_0^{\beta}d\tau\,\phi_\rho(\tau),
\end{gather}
where $\langle\mathcal{O}\rangle\equiv
Z^{-1}\int\mathcal{D}\phi_\rho\mathcal{D\phi_\sigma}\mathcal{O}
e^{-S}$
denotes the expectation value of an observable $\mathcal{O}$.
In Fig. \ref{fig_q_u1},
we plot $\langle Q\rangle$ for $V=4$ and $8$
as a function of the gate voltage $N$.
As the temperature decreases (the inverse temperature $L$ increases),
the Coulomb blockade is enhanced, and the maximum and minimum point
of the values of $\langle Q \rangle - eN$ approaches $N=1/2$ for each $V$.
If we focus on the lowest temperature $L = 400$,
the curve appears to converge except near the degeneracy point $N=1/2$.
Therefore, we can expect that the data at $L=400$
is comparable to the analytic result for $T=0$ in the strong tunneling limit ($V \rightarrow 0$).
In Ref.~\onlinecite{matveev2},
the charge field $\phi_\rho$ is pinned down to $\pi N$
due to the large charging energy $T\ll E_C<D$,
and the backscattering strength $V$ is renormalized
by the charge fluctuation. As a result, the average charge
is calculated for small $V$ (corresponding to strong tunneling) as
\begin{gather}
\frac{\langle Q\rangle}{e}=N+\frac{2\gamma|r|^2}{\pi^2}
\log({\rm e}\,|r|^2\cos^2\pi N)\sin 2\pi N,\label{charge}
\end{gather}
where $\gamma={\rm e}^{\rm C}$ with the Euler's constant ${\rm
C}\simeq0.5772$.
The inset of the Fig.~\ref{fig_q_u1} shows
the comparison for $V=4$ and $8$ between our data of $L=400$
and Eq.~(\ref{charge}).
For $V=4$, the analytic result shown as a solid line
is almost superposed on our data points;
the expression (\ref{charge}) based on the pinning of $\phi_\rho$
describes well the Coulomb oscillation of $\langle Q\rangle$.
For $V=8$, on the other hand, our data points away from $N=1/2$
disagree with the analytic expression;
the system is no longer in the strong tunneling regime
for such a large value of $V$.
It should be noted that,
the Coulomb blockade grows monotonically with increasing $V$ and $L$,
and that the reentrant behaviors in the strong tunneling regime
reported in the NRG results\cite{lebanon} are not observed in our simulation.

\begin{figure}[t]
\begin{center}
\includegraphics[width=.9\figurewidth]{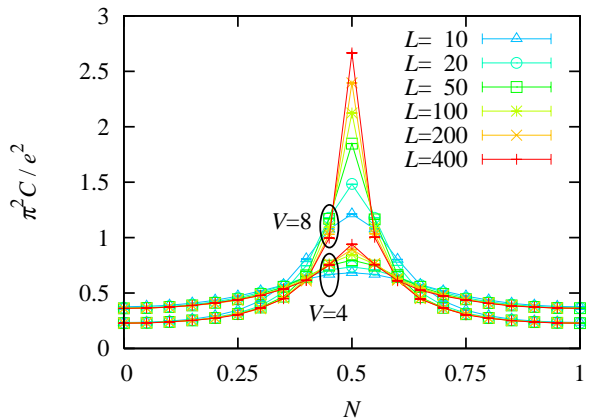}
\caption{\label{fig_c_u1}(Color online) Coulomb oscillation of the capacitance
of the dot for different temperatures $L= 10$, $20$, $50$, $100$, $200$,
and $400$.
A bundle of curves with smaller amplitude corresponds to $V=4$,
and the other to $V=8$.
}
\end{center}
\end{figure}

Near $N=1/2$ the oscillation curve in Fig. \ref{fig_q_u1}
slightly depends on $L$ at lower temperatures.
This is related to
the singularity derived from the two-channel Kondo problem
hidden in the Coulomb blockade system at the degeneracy point.
To see the peculiar nature of the system more clearly,
we now measure the capacitance of the dot defined as
\begin{gather}
C=-\frac{\partial}{\partial V_g}\langle Q\rangle=\frac{e^2}{\pi^2}\beta[\langle\bar{\phi}_\rho{}^2\rangle
-\langle\bar{\phi}_\rho\rangle^2].
\end{gather}
In Fig. \ref{fig_c_u1},
we plot the capacitance for $V=4$ and $8$ as a function of the gate voltage $N$
at low temperatures.
The amplitude of the Coulomb oscillation of the capacitance grows larger
as $L$ and $V$ increase.
The temperature dependence strongly depends on the gate voltage;
away from $N=1/2$, the capacitance is convergent at low temperatures
whereas the peak at $N=1/2$ keeps growing and sharpening
up to the lowest temperature $L = 400$.

\begin{figure}[tb]
\begin{center}
\includegraphics[width=.9\figurewidth]{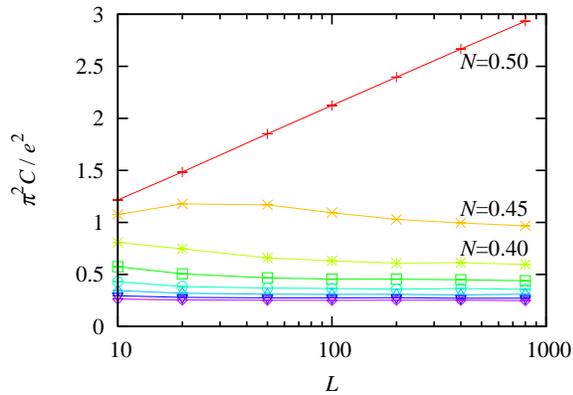}
\caption{\label{fig_c_g}(Color online)
Temperature dependence of the capacitance of the dot for $V=8$.
The gate voltage $N$ is varied from $0.15$ to $0.50$ (from bottom to top)
at intervals of $0.05$.
}
\end{center}
\end{figure}

In Fig.~\ref{fig_c_g},
we show the capacitance for intermediate tunneling $V=8$ as a function of the inverse temperature $L$
for different values of the gate voltage $N$.
In this figure, the temperature dependence of the capacitance is classified into
three different types: when the temperature decreases,
(I) for $N\le 0.40$, the capacitance decreases monotonically;
(II) for $N=0.45$, the capacitance once increases, and
decreases below a crossover temperature;
(III) for $N=0.50$, the capacitance increases monotonically.
In Ref.~\onlinecite{matveev2},
it is suggested from analytic results that the low-energy property of the Coulomb blockade system
near the degeneracy point ($N=1/2$)
is governed by the physics of the two-channel Kondo problem
for an {\it arbitrary} strength of tunneling.
This prediction is supported by
the logarithmic divergence of the capacitance for the type III.
The transient increase in the capacitance for the type II
is also a sign of the two-channel Kondo physics.
For $N\ne 1/2$, an intrinsic low energy cutoff in the strong tunneling limit 
($V \rightarrow 0$) is given by\cite{matveev2}
\begin{gather}
\Gamma=\frac{8\gamma|r|^2}{\pi^2}E_C\cos^2\pi N.
\end{gather}
At low temperatures below $\Gamma$, the logarithmic divergence
characteristic of the two-channel Kondo model disappears.
One can confirm in Fig. \ref{fig_c_g} that the crossover temperature
in the type II is roughly given by $\Gamma$.

\begin{figure}[tb]
\begin{center}
\includegraphics[width=.9\figurewidth]{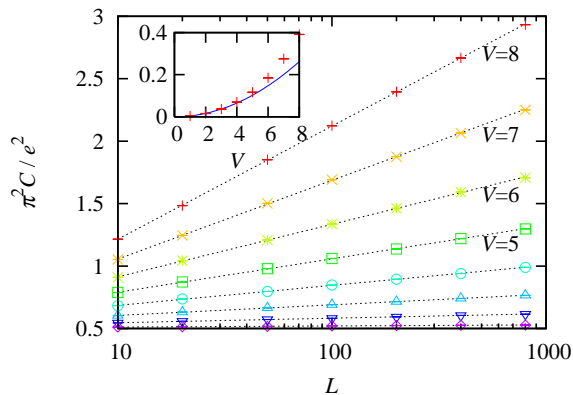}
\caption{\label{fig_c_g05}(Color online) Logarithmic divergence of the
capacitance of the dot for $N=1/2$.
The capacitance is plotted as a function of the inverse temperature $L$.
The logarithmic fit of the data points for each value of $V$ is also shown by a dotted line.
Inset: $V$-depenence of the coefficient of the logarithmic divergence.
The solid line shows the analytic result.
}
\end{center}
\end{figure}

In order to study in detail the logarithmic divergence of the capacitance,
we focus on the degeneracy point $N=1/2$
and show the capacitance in Fig.~\ref{fig_c_g05}
as a function of the inverse temperature $L$.
The results for different backscattering strengths
up to $V=8$ are shown.
One can see that our data points diverge logarithmically
with decreasing temperature for any value of $V$
as far as our simulation is performed.
Finally, we compare this logarithmic behavior of our data
with the analytic result obtained in the strong tunneling limit $(V\rightarrow 0)$.
According to the analytic expression,\cite{matveev2}
the capacitance diverges at low temperatures as
\begin{gather}
C=C_0\biggl[1+\frac{4\gamma|r|^2}{\pi}\log\biggl(\frac{E_C}{T}\biggr)\biggr].
\label{capacitance}
\end{gather}
We can directly compare the coefficient of the logarithm
between Fig.~\ref{fig_c_g05} and Eq.~(\ref{capacitance}).
In the inset of Fig. \ref{fig_c_g05},
the coefficient derived from the logarithmic fit is plotted as a function of $V$.
For small $V$, i.e., in the strong tunneling regime,
our data points agree well with the analytic result
$4\pi\gamma|r|^2C_0/e^2\simeq 0.004429V^2$ shown as a solid line.
For larger values of $V$, however, 
they grows faster than the analytic result.
Thus, the analytic expression (\ref{capacitance}) fails
for intermediate strength of tunneling.
Nevertheless, the logarithmic divergence in Fig.~\ref{fig_c_g05}
in this regime clearly indicates that the present system is understood by 
the physics of the two-channel Kondo model for any value of $V$.

In summary, we have started with the bosonized Hamiltonian of 
an open quantum dot, and demonstrated the PIMC simulation of 
the Coulomb blockade phenomena at low temperatures.
The Coulomb oscillation of the average charge and capacitance of the dot
is studied in the intermediate tunneling regime, where the analytical result
for strong tunneling is invalid.  
The amplitude of oscillation monotonically increases
with decreasing temperature or increasing the backscattering strength.
We confirmed that logarithmic divergence of the capacitance 
appears at the degeneracy point $N=1/2$ even in the intermediate tunneling regime.
This observation indicates that the present system has a two-channel Kondo nature
for an arbitary strength of tunneling.

We are grateful to Y. Utsumi for helpful discussion.
The computation in this work has been done using the facilities
of the Supercomputer Center, Institute for Solid State Physics,
University of Tokyo.

\end{document}